# Controlling magnetic exchange and anisotropy by non-magnetic ligand substitution in layered $M$P$X_3$ ($M$ = Ni, Mn; $X$ = S, Se)

Rabindra Basnet[1], K. Kotur[2], M. Rybak[3], Cory Stephenson[1], Samuel Bishop[4#], C. Autieri[5,6†], M. Birowska[2‡], Jin Hu[1,7*]

[1]*Department of Physics, University of Arkansas, Fayetteville, Arkansas 72701, USA*

[2]*Faculty of Physics, University of Warsaw, Pasteura 5, PL-02093 Warsaw, Poland*

[3]*Department of Semiconductor Materials Engineering, Faculty of Fundamental Problems of Technology, Wroclaw University of Science and Technology, Wybrzeże Wyspiańskiego 27, PL-50370 Wroclaw, Poland*

[4]*Haas Hall Academy, Fayetteville, Arkansas 72701, USA*

[5]*International Research Centre Magtop, Institute of Physics, Polish Academy of Sciences, Aleja Lotników 32/46, PL-02668 Warsaw, Poland*

[6]*Consiglio Nazionale delle Ricerche CNR-SPIN, UOS Salerno, I-84084 Fisciano (SA), Italy*

[7]*Institute for Nanoscience and Engineering, University of Arkansas, Fayetteville, Arkansas 72701, USA*

## Abstract

Recent discoveries in two-dimensional (2D) magnetism have intensified the investigation of van der Waals (vdW) magnetic materials and further improved our ability to tune their magnetic properties. Tunable magnetism has been widely studied in antiferromagnetic metal thiophosphates $M$P$X_3$. Substitution of metal ions $M$ has been adopted as an important technique to engineer the magnetism in $M$P$X_3$. In this work, we have studied the previously unexplored chalcogen $X$ substitutions in $M$P$X_3$ ($M$ = Mn/Ni; $X$ = S/Se). We synthesized the single crystals of MnPS$_{3-x}$Se$_x$ ($0 \leq x \leq 3$) and NiPS$_{3-x}$Se$_x$ ($0 \leq x \leq 1.3$) and investigated the systematic evolution of the magnetism with varying $x$. Our study reveals the effective tuning of magnetic interactions and anisotropies in both MnPS$_3$ and NiPS$_3$ upon Se substitution. Such efficient engineering of the magnetism provides a suitable platform to understand the low-dimensional magnetism and develop future magnetic devices.

[†]autieri@magtop.ifpan.edu.pl; [‡]birowska@fuw.edu.pl; [*]jinhu@uark.edu

[#]Current address: Williams College, Massachusetts 01267, USA

1.  **Introduction**

    The recently discovered two-dimensional (2D) magnetic materials have attracted intensive attention because of the possible new phenomena arising from 2D magnetism and the promising potential for spintronic applications [1–6]. The integration of 2D magnets in nanodevices and heterostructures has further fueled the investigation of their magnetic and electronic properties, thus offering a fertile platform to design next-generation devices [7–19]. Such developments in 2D magnetism have motivated the search for novel magnetic van der Waals (vdW) materials to explore the low-dimensional magnetism in real materials.

    One interesting class of magnetic vdW materials is antiferromagnetic (AFM) $M$P$X_3$ ($M$ = transition metal ions, $X$ = chalcogen ions), in which the transition metal ions carry localized magnetic moments in a layered honeycomb lattice [20–23]. Their magnetic properties are strongly influenced by the transition metal $M$. The magnetic exchange and anisotropy are found to vary with the choice of $M$ [23–40]. Such tuning of magnetism by altering $M$ has led to the study of a series of polymetallic "mixed" $M$P$X_3$ compounds such as $Ni_{1-x}Mn_xPS_3$, $Mn_xFe_{1-x}PS_3$, $Fe_{1-x}Ni_xPS_3$, $Mn_{1-x}Zn_xPS_3$, $Ni_{1-x}Co_xPS_3$ and $Mn_{1-x}Fe_xPSe_3$ [34,37,39,41–53]. Tunable magnetism arising from the interplay between competing magnetic interactions, magnetic anisotropy, and spin fluctuations has been observed in these mixed compounds [34,37,39,41–52], providing promising candidates to explore novel phenomena originating from 2D magnetism.

    Compared with the metal substitution in the $M$ site [34,37,39,41–53], chalcogen substitution in the $X$ site leaves the magnetic ions intact, hence offering a relatively clean approach to modify the magnetic exchange interactions. The effectiveness of substituting non-magnetic ligand atoms to engineer the magnetism in vdW magnets has been demonstrated recently [54,55]. For example, in chromium halide, varying the ratio of halides is found to

effectively control the ordering temperature and magnetic anisotropy [54]. Furthermore, the competing spin-orbit coupling strength of Cr and halides leads to a new frustrated regime and modified interlayer coupling in $CrCl_{3-x-y}Br_xI_y$ [55]. However, the chalcogen-substitution effects on magnetism remain elusive in $MPX_3$, with only a few syntheses and structural characterization works have been reported [56–59].

In this work, we conducted a systematic study on the magnetic properties of $MnPS_{3-x}Se_x$ and $NiPS_{3-x}Se_x$ ($0 \leq x \leq 3$). We found very different doping dependences for Néel temperature ($T_N$) in those two material systems, likely due to the difference in their dominant exchange interactions i.e, direct $M$-$M$ interaction in $MnPS_3$ whereas superexchange $M$-$X$-$M$ interaction in $NiPS_3$. Furthermore, chalcogen substitution also effectively controls the magnetic anisotropy, which is manifested by the efficient tuning of the magnetic easy axis and spin flop (SF) transition. Such tunable magnetism achieved by non-magnetic substitutions offers a useful technique to engineer low-dimensional magnetism and provides further insights for the development of magnetic materials-based nanodevices.

## 2. Experiment

The $MnPS_{3-x}Se_x$ and $NiPS_{3-x}Se_x$ single crystals used in this work were synthesized by a chemical vapor transport method using $I_2$ as the transport agent. Elemental powders with desired ratios were sealed in a quartz tube and heated in a two-zone furnace for a week. The $MnPS_{3-x}Se_x$ single crystals were grown with a temperature gradient from 650 to 600 °C, whereas 750 to 550 °C was used for $NiPS_{3-x}Se_x$ growth. The polycrystalline $NiPS_{3-x}Se_x$ samples used in this work were grown using a self-flux method at 750 °C. The as-grown polycrystalline samples were annealed at 750 °C for 4 days to minimize the possible impurity phase. Such annealing process is necessary to obtain a pure phase for characterizing magnetic properties. The elemental compositions and crystal structures of the obtained crystals were examined by energy-dispersive x-ray spectroscopy (EDS) and x-ray diffraction (XRD), respectively. Magnetization measurements were performed in a physical property measurement system (PPMS, Quantum Design).

The calculations were performed in the framework of the DFT+U [60] approach as implemented in VASP software [61,62]. The effective on-site Coulomb exchange parameters were set to $U = 5$ eV and $U = 6$ eV for $3d$ states of Mn and Ni atoms, respectively. A cutoff of 400 eV was chosen for the plane-wave basis set and a **k**-mesh of 10×6×2 and 10×6×9 was taken to sample the first Brillouin zone on Γ-centered symmetry reduced Monkhorst-Pack mesh for monolayer and bulk systems, respectively. The denser **k**-mesh grids equal to 15×9×13 was taken with spin-orbit coupling (SOC) included. The standard exchange-correlation functionals neglect the non-local nature of dispersive forces, which are crucial for layered materials and adsorption molecules on the surfaces [63–65]. Thus, the semi-empirical Grimme method was applied [66]. The lattice parameters have been fixed to the experimental ones. The positions of the atoms were relaxed until the maximal force per atom was less than $10^{-3}$ eV/Å. The noncollinear magnetism and SOC were included in our calculations.

3. **Results and discussion**

In each layer of *MPX*$_3$, the transition metals *M* are surrounded by P$_2$X$_6$ clusters as shown in Fig. 1a. So, 0the chalcogen substitution in the *X* site would modify the local environment of $M^{2+}$ within honeycomb layers. Our extensive crystal growth efforts have resulted in sizable single crystals of MnPS$_{3-x}$Se$_x$ with *x* up to 3 (i.e., full replacement of S by Se). As shown in Fig.

1b, these crystals are relatively transparent, showing a gradual color change from green ($x = 0$) to wine red ($x = 3$), which indicates the variation of the optical gap. On the other hand, for NiPS$_{3-x}$Se$_x$, good single crystals can only be obtained for $x$ up to 1.3 (Fig. 1c). Such difficulty in growing single crystals for Se-rich samples could be the reason for the very limited studies on NiPSe$_3$ [67] as compared to NiPS$_3$. The successful substitution in both MnPS$_{3-x}$Se$_x$ and NiPS$_{3-x}$Se$_x$ is demonstrated by the composition analyses using EDS. Furthermore, As shown in Figs. 1b and 1c, the (00$l$) XRD peaks show systematic low-angle shift with increasing the Se content, consistent with the elongation of $c$-axis when smaller S is replaced by larger Se.

To investigate the effects of Se substitution on magnetic properties, we measured the temperature dependence of molar susceptibility ($\chi$) for MnPS$_{3-x}$Se$_x$ and NiPS$_{3-x}$Se$_x$ under the out of plane ($H \perp ab$) (red color) and in-plane ($H // ab$) (blue color) magnetic fields (Figs. 2a and 3a). To obtain the precise transition temperature for each sample, we used the peak position of the derivative $d\chi/dT$ to define $T_N$ (Figs. 2b and 3b), which has also been widely used in previous studies [39,51,53,68]. As shown in Figs. 2a and 3a, the out-of-plane ($\chi_\perp$) and in-plane ($\chi_{//}$) susceptibility for both MnPS$_{3-x}$Se$_x$ and NiPS$_{3-x}$Se$_x$ overlap in the paramagnetic (PM) state but start to deviate below $T_N$ (denoted by black triangles). In earlier studies, both substantial [24,39,53,69] and the lack [25,33,35,51,68] of magnetic anisotropy in the PM state have been reported. It has also been pointed out that the sample holder may contribute to the observed magnetic anisotropy, especially for the NiPS$_3$ samples whose magnetization is relatively weak [25]. Therefore, we have been very careful in the magnetization measurements, for which we used the same quartz sample holder for both $\chi_\perp$ and $\chi_{//}$ measurements. For Se-rich NiPS$_{3-x}$Se$_x$ ($x = 2$ and 3) (Fig. 3a), $T_N$ is obtained from the measurements on polycrystalline samples so that $\chi_\perp$ and $\chi_{//}$ cannot be obtained. The extracted $T_N$ for the end compounds ($x = 0$ or

3) MnPS$_3$, MnPSe$_3$, NiPS$_3$, and NiPSe$_3$ are 78.5, 74, 155, and 212 K, respectively, consistent with the previous studies [23,25,27,29,32,50,67,70,71].

Substituting S by Se leads to systematic variations of $T_N$ in both material systems. Such $T_N$ evolution is completely different from that caused by metal substitutions in MnPS$_3$ and NiPS$_3$ [34,37,39,46]. In polymetallic mixed compounds such as Ni$_{1-x}$Mn$_x$PS$_3$ [37,39], Mn$_x$Fe$_{1-x}$PS$_3$ [46], and Mn$_{1-x}$Zn$_x$PS$_3$ [34], $T_N$ is drastically reduced by substituting magnetic or non-magnetic metal ions, reaching minimum around $x = 0.5$. However, chalcogen substitution only slightly reduces $T_N$ in MnPS$_{3-x}$Se$_x$, from 78.5 K for MnPS$_3$ to 74 K for MnPSe$_3$ (Figs. 2a and 2c). For NiPS$_{3-x}$Se$_x$, Se substitution causes $T_N$ to increase monotonically (Figs. 3a and 3c), which is distinct from the sharp decrease in metal substituted NiPS$_3$ [37,39]. As will be discussed below, the observed evolutions of $T_N$ with chalcogen substitutions in these two material systems can be understood in terms of the anion-mediated superexchange interactions in addition to the direct *M-M* exchange. The distinct doping dependences of $T_N$ in those compounds can be attributed to their different magnetic interactions [72,73].

In *MPX*$_3$ compounds, substituting S by larger Se expands the in-plane lattice [59,74] and leads to the attenuation of the direct *M-M* interaction within the metal ion plane [59,65]. In MnPS$_3$, the neighboring Mn moments are found to be antiparallel (Fig. 2d) [72,73] and the magnetism in this compound is known to be governed by the Mn-Mn direct exchange interaction [67,72,73]. Therefore, the systematic suppression of $T_N$ by Se substitution in MnPS$_{3-x}$Se$_x$ can be attributed to the weakened direct exchange interaction between the nearest-neighbor Mn ions, due to the elongated Mn-Mn bond. Indeed, weaker Mn-Mn exchange in *M*PSe$_3$ in comparison to *M*PS$_3$ has been theoretically proposed [67] and experimentally demonstrated by neutron scattering experiment [70].

To clarify the evolution of magnetic exchange interactions (*J*) upon chalcogen substitutions, we performed the DFT calculations. In order to evaluate the Néel temperatures of bulk MnPS$_3$, MnPSe$_3$, NiPS$_3$ and NiPSe$_3$ structures, we first examine the various magnetic ordering such as antiferromagnetic ones: Néel (AFM-N), zigzag (AFM-z), stripy (AFM-s), and ferromagnetic (FM) one, as reported in another work [76]. We consider the in-plane and out-of-plane directions of the spins, following the previous report [77]. Namely, we consider the Heisenberg Hamiltonian with a single ion anisotropy A:

$$H = E_0 - \frac{1}{2}\sum_{ij} J_{ij}\bar{S}_i\bar{S}_j - \frac{1}{2}\sum_{ij} \lambda_{ij} S_i^z S_j^z - A\sum_i (S|i^z)^2 \tag{1}$$

where ½ accounts for the double-counting, $E_0$ denotes the energy of the nonmagnetic system, $S_i$ is a spin magnetic moment of the atomic site *i*. $J_{ij}$ and $\lambda_{ij}$ are the isotropic and anisotropic exchange couplings between the atomic site *i* and *j*, respectively. The off-diagonal isotropic exchange terms have been neglected. The details of these calculations along with derived equations are presented in Supporting Information (SI). Here, we consider the magnetic exchanges in the monolayer systems up to the third nearest neighbors, neglecting the exchange coupling from adjacent layers. Finally, the Néel temperature has been evaluated in the mean-field approach [72], which takes the following form

$$T_N^{Mn} = S(S+1)(-3J_1 - 6J_2 - 3J_3)/(3k_B) \tag{2}$$

$$T_N = S(S+1)(J_1 - 2J_2 - 3J_3)/(3k_B) \tag{3}$$

for the Mn and Ni compounds, respectively. The difference between equations (2) and (3) is attributed to the different magnetic order of these two compounds. Note that the different sign of $J_1$ in the equations (2) and (3) produces a different dependence of the $T_N$ from $J_1$. Where spin S is equal to 5/2 and 1 for Mn and Ni, respectively and $k_B$ is the Boltzmann constant. Similar expressions have been obtained in an earlier work [72]. The results are collected in Table 1 and are in excellent agreement with previous studies [72,73,78]. Note that the Néel temperatures obtained here are overestimated, which is a well-known fact for the systems which exhibit strong critical fluctuations. Although, these critical values are overestimated as expected from the mean-field approximation, the change between the Se and S systems reflects qualitatively the change of the critical temperatures whenever the S atoms are substituted by Se. As shown in Table 1, the calculated nearest-neighbor interaction ($J_1$) and third nearest-neighbor interaction ($J_3$) are reduced upon replacing Se for S in MnP$X_3$, suggesting the suppression of in-plane exchange interaction on Se substitution, which is in line with the neutron scattering experiment [70]. Note, that for both MnPSe$_3$ and NiPSe$_3$ the bond angle between the M-S-M atoms is closer to 90° in comparison to their corresponding Sulphur structures, see Table S1, pointing to the enhancement of the nearest neighbor FM superexchange according to Goodenough-Kanamori-Anderson rules [79,80]. Here, we did not comment the changes in the second nearest-neighbor interaction ($J_2$) because it is negligible in both Mn and Ni systems as reported in Table 1. Such attenuation of in-plane magnetic interactions can explain the experimental observation of declining $T_N$ with Se substitution in MnPS$_{3-x}$Se$_x$.

The situation is different in Se-substituted NiPS$_3$. Earlier neutron scattering experiments [67,72,73] have demonstrated that the magnetic interactions in NiPS$_3$ occur only through a superexchange pathway. The direct exchange among the metal ions, however, does not exist because of the filled $t_{2g}$ orbitals for Ni$^{2+}$ [67,72,73]. Therefore, though direct exchange usually significantly influences $J_1$ in magnetic materials, $J_1$ in NiPS$_3$ is superexchange in nature [72,73] and weakly dependent on the Ni-Ni distance [75]. The observed systematic increase of $T_N$ in Se substituted NiPS$_3$ should be ascribed to the enhanced superexchange interactions. Substituting the non-magnetic ligand atoms is known to effectively tune superexchange in various materials [54,75,81]. In general, replacing a smaller ligand with a larger one usually leads to enhanced superexchange interaction because of the stronger orbital overlap due to greater atom orbitals [54,75]. This can be seen in our DFT calculation, which demonstrates stronger superexchange $J_1$ and $J_3$ for NiPSe$_3$ in comparison to NiPS$_3$ (Table 1). The result of the calculation depicts the FM and AFM nature of $J_1$ and $J_3$ respectively in the Ni system, which is consistent with the reported magnetic structure of NiPS$_3$ [25] shown in Fig. 3(d). The $J_1$ of both Mn and Ni systems become more ferromagnetic, but this has a different effect on the two systems due to the different magnetic order as we can see from equation (2) and (3). Although both FM $J_1$ and AFM $J_3$ are enhanced upon Se substitution, the larger magnitude of $J_3$ (almost 4 times) than $J_1$ explains the stronger AFM interaction in NiPSe$_3$ than NiPS$_3$ proposed in an earlier study [67]. This also agrees well with the systematic increase of $T_N$ with Se substitution in NiPS$_{3-x}$Se$_x$. Similar $T_N$ enhancement by substituting with larger ligand atoms has also been observed in many other compounds whose magnetic interactions are mainly mediated by superexchange couplings, such as CrCl$_{3-x}$Br$_x$ [54] and CuCr$_{1.5}$Sb$_{0.5}$S$_{4-x}$Se$_x$ [81].

Though the elongation of the Ni bonds may not strongly affect exchange interactions in NiPS$_{3-x}$Se$_x$ as mentioned above, it may still mediate the in-plane magnetic interactions. As shown in Fig. 3(d), within the Ni layer, the Ni moments form a bi-collinear AFM order consisting of ferromagnetic (FM) chains along the $a$-axis [72,73]. Therefore, the expansion of the in-plane lattice would weaken the intra-chain FM couplings, which may favor the overall AFM interactions of the sample.

Though the different Se substitution-dependent ordering temperatures in MnPS$_{3-x}$Se$_x$ and NiPS$_{3-x}$Se$_x$ can be understood in terms of the mediation of magnetic exchange interactions, the very different effectiveness of tuning $T_N$ by substitution is remaining. As shown in Figs. 2c and 3c, $T_N$ slightly decreases only by 6% from 78.5 K for MnPS$_3$ to 74 K in MnPSe$_3$, but increases remarkably in NiPS$_{3-x}$Se$_x$, from 155 K for NiPS$_3$ to 212 K in NiPSe$_3$. Such stronger composition dependence in NiPS$_{3-x}$Se$_x$ in contrast to MnPS$_{3-x}$Se$_x$ can be attributed to the larger magnitudes of exchange couplings $J_i$ for Ni structures in comparison to their Mn counterparts (Table 1). Moreover, as suggested by the previous neutron scattering experiment [60], MnPSe$_3$ shows stronger interlayer exchange interaction $J_c$ than MnPS$_3$ which may stabilize the magnetic order, so the enhanced $J_c$ with Se substitution may also offset the decrease of $T_N$ driven by reduced intra-layer exchange interaction, leading to the observed weak composition dependence in MnPS$_{3-x}$Se$_x$.

It is worthwhile to compare the distinct effects between chalcogen and metal substitutions in *M*P*X*$_3$ compounds. For polymetallic *M*P*X*$_3$ compounds in which the magnetic *M* is substituted by other magnetic or non-magnetic metal elements, a remarkable reduction in $T_N$ has been observed [34,37,39,46,50], which has been ascribed to the suppression of magnetic interactions by random distributions of mixed metal ions, as well as the magnetic frustrations when substituting with magnetic metal elements. Indeed, recent DFT calculation has revealed frustrations among Ni and Mn atoms in Ni$_{0.75}$Mn$_{0.25}$PS$_3$ resulting from the competing Néel and zig-zag AFM configurations [78]. The light suppression in $T_N$ in chalcogen-substituted *M*P*X*$_3$ may imply weaker frustrations compared to the case of metal ion substitutions. This is also consistent with the crystal structures of *M*P*X*$_3$ compounds in which the chalcogen atoms are located away from the magnetic layers (Fig. 1a). Therefore, substituting chalcogen layers mainly modifies the environment above and below the magnetic layers, rather than inducing strong magnetic impurities and magnetic dilutions for magnetic [37,39,46] and non-magnetic [34,43] metal-ion substitutions respectively. Hence the chalcogen substitutions could be a better approach to modify magnetism without strongly destabilizing the magnetic orderings in $M$P$X_3$.

Metal-ion substitutions in NiPS$_3$ and MnPS$_3$ have also been found to be effective in tuning magnetic anisotropies [37,39,50,52]. For example, varying the Ni:Mn ratio in Ni$_{1-x}$Mn$_x$PS$_3$ can re-orientate the magnetic easy axes from nearly within the *ab*-plane to along the *c*-axis [37,39]. In NiPS$_3$, Fe substitution can trigger a crossover from *XY* to Ising anisotropy [51,53]. Moreover, the magnetic anisotropy in MnPS$_3$ can be reduced with magnetic dilution by substituting non-magnetic Zn [34,43]. In this work, we further studied the evolution of magnetic anisotropy with chalcogen substitutions. Fig. 4a presents the isothermal field-dependent magnetization $M(H)$ at 2K for MnPS$_{3-x}$Se$_x$ ($0 \leq x \leq 3$) measured under out-of-plane ($H \perp ab$) and in-plane ($H // ab$) magnetic fields. The evolution of magnetic anisotropy can be extracted from the low field (i.e., below the critical field of spin-flop transition) magnetization. The low field magnetic susceptibility is smaller with $H \perp ab$ in Se-less samples. Increasing Se content *x* gradually reduces anisotropy, and eventually leads the low field out-of-plane magnetization to surpass the in-plane one for $x > 0.7$. Such observation is expected since the two end compounds MnPS$_3$ and MnPSe$_3$ have different easy axis [36,43,50,70,71,82]. As shown in Figs. 4c and 4d, the Mn moments in MnPS$_3$ are aligned along the out-of-plane direction [36,43], whereas they mostly lie within the basal plane for MnPSe$_3$ [50,70,71,82]. The rotation of the easy axis with Se substitution is also supported by the temperature dependent susceptibility below $T_N$ (Fig. 2a), in which $\chi_\perp$ is smaller than $\chi_{//}$ for samples with *x* up to 0.7 but larger for samples with more Se.

The moment orientations in $M$P$X_3$ compounds are resulting from the competition between the dipolar and single-ion anisotropy, which favor an out-of-plane and an in-plane moment orientation, respectively. The single-ion anisotropy arises from the combined effects of the trigonal distortion of $MX_6$ octahedra and spin-orbit coupling (SOC) [24,37]. In MnPS$_3$, the trigonal distortion and spin-orbit splitting are negligible, so its magnetism is mainly governed by the dipolar anisotropy which results in an out-of-plane moment orientation [24]. Substituting S with Se does not significantly modify trigonal distortion [70], but it causes increased ligand SOC contribution which may enhance single-ion anisotropy [71], leading to the observed moment rotation and the switching of magnetic anisotropy when $x > 0.7$. This is supported by our DFT calculation, which shows the enhancement of single-ion anisotropy ($A$) in MnPSe$_3$ by almost one order of magnitude as compared to MnPS$_3$ (Table 1). Although the SOC which is responsible for magnetocrystalline anisotropy is accounted in our calculations, the magnetic dipolar interactions are not included in our results. Note that the very recent study has reported a newly developed exchange-correlation functional [83] which accounts for the magnetic dipole-dipole interactions. Considering the MnPS$_3$ system, the inclusion of the SOC causes the spins to lie within the basal plane (Fig. 5a). This result is obviously in contradiction to the experimental result (Fig. 4c), due to the lack of the inclusion of the magnetic dipolar anisotropy in our calculations, which might align the spins to the out-of-plane direction. In addition to enhanced $A$, it is also plausible that long-range magnetic dipolar interactions are weakened in MnPSe$_3$ than MnPS$_3$, due to the longer lattice parameters for the Se than S compounds. Thus, the dominant $A$ in Se-rich compounds can be ascribed to the switching of anisotropy from out-of-plane to in-plane direction. Furthermore, unlike MnPS$_3$ whose interlayer magnetic exchange interaction $J_c$ is negligible, a recent neutron scattering experiment has revealed considerable $J_c$ when S is replaced by Se [60], which is not accounted in our theoretical results, and will be studied elsewhere. More theoretical efforts are needed to clarify the possible coupling between $J_c$ enhancement and moment rotation.

Tuning moment orientations in $MPX_3$ modifies the SF transition [37,43,84]. MnPS$_3$ displays a SF transition when the magnetic field is applied along the out-of-plane direction, which is characterized by a drastic magnetization upturn at the SF transition field ($H_{SF}$) (Fig. 4a) [5,33–37,39]. Substituting Mn with Ni is found to strongly suppress the SF transition [37,39]. $H_{SF}$ is suppressed by half with only 5% Ni substitution, and disappears for 10% substitution, which is likely due to the reorientation of the magnetic moments when single ion anisotropy is modulated by enhanced lattice trigonal distortion upon substitution [37]. Moreover, though not as efficient as magnetic Ni substitution, the non-magnetic Zn-substitution also reduces $H_{SF}$, which has been ascribed to the weakening of magnetic anisotropy with magnetic dilution [43]. In this work, modification of the SF transition with Se substitution has also been observed. As shown in Fig. 4a, at $T = 2$ K and for $H \perp ab$, the SF transition is gradually suppressed, as manifested by the less-obvious magnetization upturn and reduced $H_{SF}$. When Se content $x$ is increased beyond 0.7, the observation is inverted. Magnetization displays linear field dependence when $H \perp ab$ but an upturn starts to appear under $H//ab$. Such spin flop transition occurs around ~2.9 T in the $x = 1.2$ sample, which remains strong for Se content up to $x = 1.3$ and is gradually suppressed with further increasing the Se content. For the end compound MnPSe$_3$ ($x = 3$), a small upturn around $\mu_0 H = 0.7$ T can be seen in the in-plane magnetization (Fig. 4a, inset).

In Fig. 4b we summarize the variation of $H_{SF}$ with Se content for both out-of-plane and in-plane field, from which the two types of SF transitions under $H \perp ab$ and $H//ab$ can be clearly seen. Generally, a SF transition in an AFM compound is characterized by the moment re-orientation that is driven by the magnetic field component parallel to the magnetic easy axis. The change of the moment orientation can be attributed to the enhanced single-ion anisotropy upon Se substitution as discussed above, which usually favors the in-plane moment orientation.

Therefore, in MnPS$_3$ whose easy axis is out-of-plane, the SF transition is characterized by Mn moment rotation toward the *ab*-plane under $H\perp ab$ [36,37]. The corresponding $H_{SF}$ is suppressed by Se substitution, because the easy axis in the substituted samples already rotates away from the out-of-plane direction.

One interesting observation is that the switching of the easy axis occurs occurring closer to MnPS$_3$ side i.e., between $x = 0.7$ and 1.2 [Fig. 4a and b], which implies the magnetism in MnPS$_3$ is softer than MnPSe$_3$. This is also consistent with our theoretical calculations. The energy difference between the in-plane and out-of-plane directions of the spins is of the order of a few hundredths of meV in MnPS$_3$ (Fig. 5a, Left panel) and can be considered negligible ($A = -0.005$ meV, Table 1), as reported in earlier works [24,72,73]. On the other hand, the energy difference in MnPSe$_3$ (Fig. 5b, Left panel) is one order of magnitude greater than in the case of MnPS$_3$, which is reflected in the enhancement of the monocrystalline anisotropy ($A = -0.037$ meV, Table 1).

Another interesting feature is that the magnetic ordering temperature does not change remarkably (Fig. 2c) when $H_{SF}$ is drastically suppressed above $x = 0.7$ (Fig. 4c). Similar observations have also been reported for Ni-substituted MnPS$_3$, in which the 10% Ni substitution can fully suppress the SF transition but leaves the ordering temperature essentially unchanged [37]. Such distinct composition dependences for $H_{SF}$ and $T_N$ have been attributed to the dominant role of the single ion anisotropy rather than the magnetic exchange in modulating SF transition [37], which may also be applicable for MnPS$_{3-x}$Se$_x$ studied in this work.

Similarly, the rise of SF transition for *H//ab* is also in line with the rotation of the easy axis toward the basal plane. For the end compound MnPSe$_3$, to the best of our knowledge, the isothermal field dependent magnetization has not been reported, though this material has been known for a long time and extensively studied [50,70,71,82]. The observed weak low-field magnetization upturn is suggestive to a SF transition under in-plane magnetic field which is in line with the in-plane Mn moment orientation (Fig. 4d). If the SF transition is real, the small $\mu_0 H_{SF} \approx 1$ T would be the lowest SF field reported so far in $MPX_3$ family [5,33–38]. Such lower $H_{SF}$ in MnPSe$_3$ might be attributed to the distinct moment re-orientation during the SF transition. For example, the moment may rotate within the basal plane so that $H_{SF}$ is lower, unlike MnPS$_3$ whose larger $H_{SF}$ may be related to the higher field required to overcome the anisotropy difference between its easy axis and the basal plane. Similar SF mechanism has also been proposed in another Mn system [85].

Now we turn our discussion to NiPS$_{3-x}$Se$_x$. As shown in Fig. 6a, consistent with the earlier magnetization studies [37,39], a metamagnetic spin flop transition under an in-plane field of ~6 T can be seen in isothermal magnetization measurements. Such observation also agrees with the nearly in-plane orientation for Ni moments [25] due to strong single-ion anisotropy in NiPS$_3$ [72,73]. Se substitution leads the SF transition to occur at a higher in-plane field of around 8 T for $x = 0.3$. Further increasing Se content leads to essentially linear field-dependent magnetization for both $H//ab$ and $H\perp ab$ up to 9 T. The increased $H_{SF}$ with Se substitution in NiPS$_3$ is in sharp contrast to observation in MnPS$_{3-x}$Se$_x$ (Fig. 4b). Unlike the MnPS$_{3-x}$Se$_x$ system for which the magnetic structure for end compounds MnPS$_3$ [36,43] and MnPSe$_3$ [50,70,71,82] have been well understood, the lack of the established magnetic structure for NiPSe$_3$ makes it difficult to clarify how Ni moment orientation may play a role in the observed substitution dependence of $H_{SF}$ in NiPS$_{3-x}$Se$_x$. Our DFT calculation has demonstrated the out-of-plane moment orientation for NiPSe$_3$ (Fig. 6b). This agrees well with the possible higher $H_{SF}$ required for Se-rich samples, in a manner similar to MnPS$_{3-x}$Se$_x$, where SF transition occurs at higher $H_{SF}$ for compounds ($x$ = 0-0.7) with the easy axis along an out-of-plane direction. In addition to the orientation of the easy axis, the stronger exchange interaction in Se-substituted NiPS$_3$ may be another possible factor that affects the $H_{SF}$. The spin-flop field at low temperatures can be approximately expressed as $H_{SF} \approx \sqrt{2H_E H_A}$ where $H_E$ and $H_A$ are effective exchange and magnetic anisotropy fields, respectively [85]. Since Se substitution enhances the exchange interactions as discussed earlier, increased $H_{SF}$ is expected. To better clarify the mechanism for the evolution of SF transition in NiPS$_{3-x}$Se$_x$, future neutron scattering experiments even on polycrystalline samples, magnetization measurements under high magnetic field, and theoretical efforts would be helpful.

In conclusion, we studied the magnetic properties of previously unreported Se-substituted MnPS$_3$ and NiPS$_3$. We found distinct tuning of $T_N$ in MnPS$_{3-x}$Se$_x$ and NiPS$_{3-x}$Se$_x$, likely attributed to different exchange interactions in pristine MnPS$_3$ and NiPS$_3$. In addition, the magnetic anisotropy is also efficiently modulated with S-Se substitutions. Our findings provide a suitable platform for a deeper understanding of low-dimensional magnetism and potential spintronics applications.

**Acknowledgment**

Work at the University of Arkansas (crystal growth and measurements) is supported by the U.S. Department of Energy, Office of Science, Basic Energy Sciences program under Grant No. DE-SC0022006. C. A. is supported by the Foundation for Polish Science through the International Research Agendas program co-financed by the European Union within the Smart Growth Operational Programme. M. B. acknowledges support within grant UMO-2016/23/D/ST3/03446

Access to computing facilities of PL-Grid Polish Infrastructure for Supporting Computational Science in the European Research Space and of the Interdisciplinary Center of Modeling (ICM), University of Warsaw is gratefully acknowledged. Financial support from the University of Warsaw under the "Excellence Initiative - Research University" project is acknowledged by M.B. We made use of computing facilities of TU Dresden ZIH within the project "TransPheMat".

**Table 1.** The exchange couplings $J_i$, the strength of a single ion anisotropy $A$ and $M$-$X$-$M$ bond angle assuming the experimental lattice parameters. Positive (negative) values of $J_i$ and $A$ indicate the FM (AFM) couplings and out-of-plane (in-plane) direction of spins, respectively. The critical temperature is evaluated in mean-field approach.

| Structure | $J_1$ (meV) | $J_2$ (meV) | $J_3$ (meV) | $\lambda_1$ (meV) | $\lambda_2$ (meV) | $\lambda_3$ (meV) | $A$ (meV) | $T_N$ (K) | ∢$M$-$X$-$M$ (°) |
|---|---|---|---|---|---|---|---|---|---|
| MnPS$_3$ ($a$ = 6.07 Å, $b$ = 10.55 Å) | -1.22 | -0.06 | -0.43 | -4×10$^{-5}$ | -4×10$^{-5}$ | 7×10$^{-5}$ | -0.005 | 181 | 83.29 |
| MnPSe$_3$ ($a = b$ = 6.32 Å) | -1.07 | -0.06 | -0.24 | -3×10$^{-3}$ | -4×10$^{-3}$ | -2×10$^{-3}$ | -0.037 | 145 | 83.34 |
| NiPS$_3$ ($a$ = 5.81 Å, $b$ = 10.07 Å) | 3.53 | 0.33 | -14.06 | 8×10$^{-4}$ | -8×10$^{-4}$ | 2×10$^{-3}$ | -0.108 | 349 | 85.13 |
| NiPSe$_3$ ($a$ = 6.15 Å, $b$ = 10.66 Å) | 4.53 | -0.13 | -16.11 | -5×10$^{-3}$ | -3×10$^{-2}$ | 9×10$^{-3}$ | 0.271 | 411 | 86.40 |

**Figure 1**

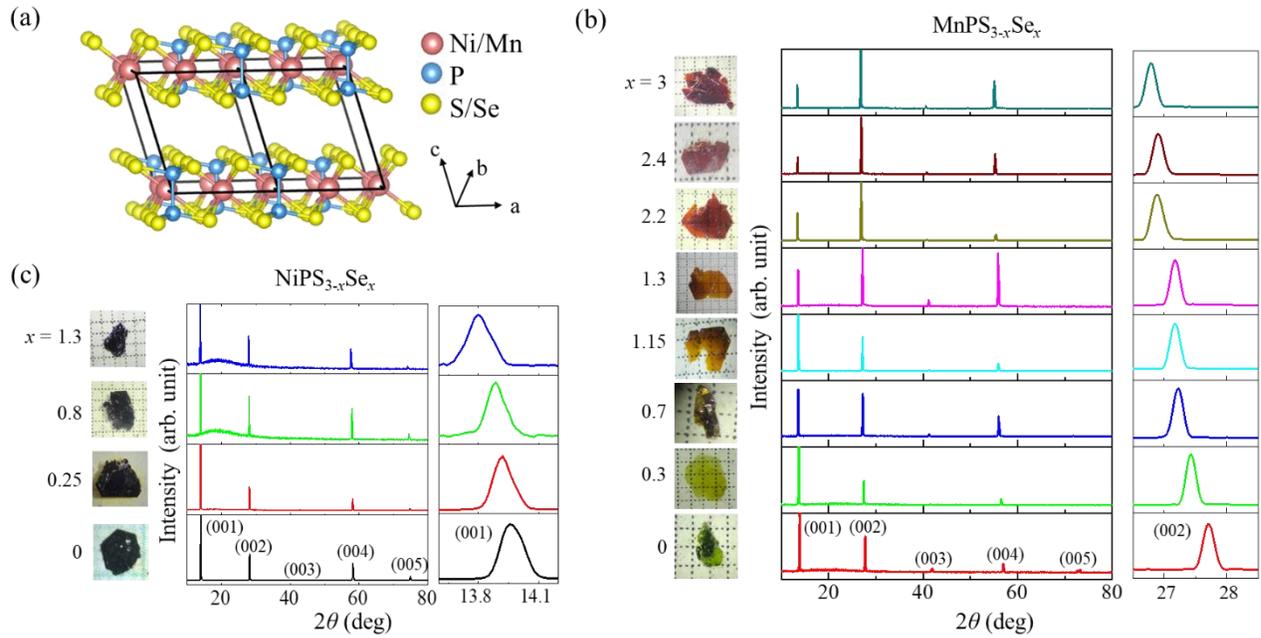

FIG. 1. (a) Crystal structure of $MPX_3$ (M = Mn/Ni; $X$ = S/Se). (b) Optical microscope images of the as-grown single crystals and Single-crystal x-ray-diffraction pattern of $MnPS_{3-x}Se_x$ ($0 \leq x \leq 3$) showing the (00$L$) reflections. Right panels show (002) diffraction peak. (c) Optical microscope images of the as-grown single crystals and Single-crystal x-ray-diffraction pattern of $NiPS_{3-x}Se_x$ ($0 \leq x \leq 1.3$) showing the (00$L$) reflections. Right panels show (001) diffraction peak. The value of Se content $x$ for each sample is determined by EDS.

**Figure 2**

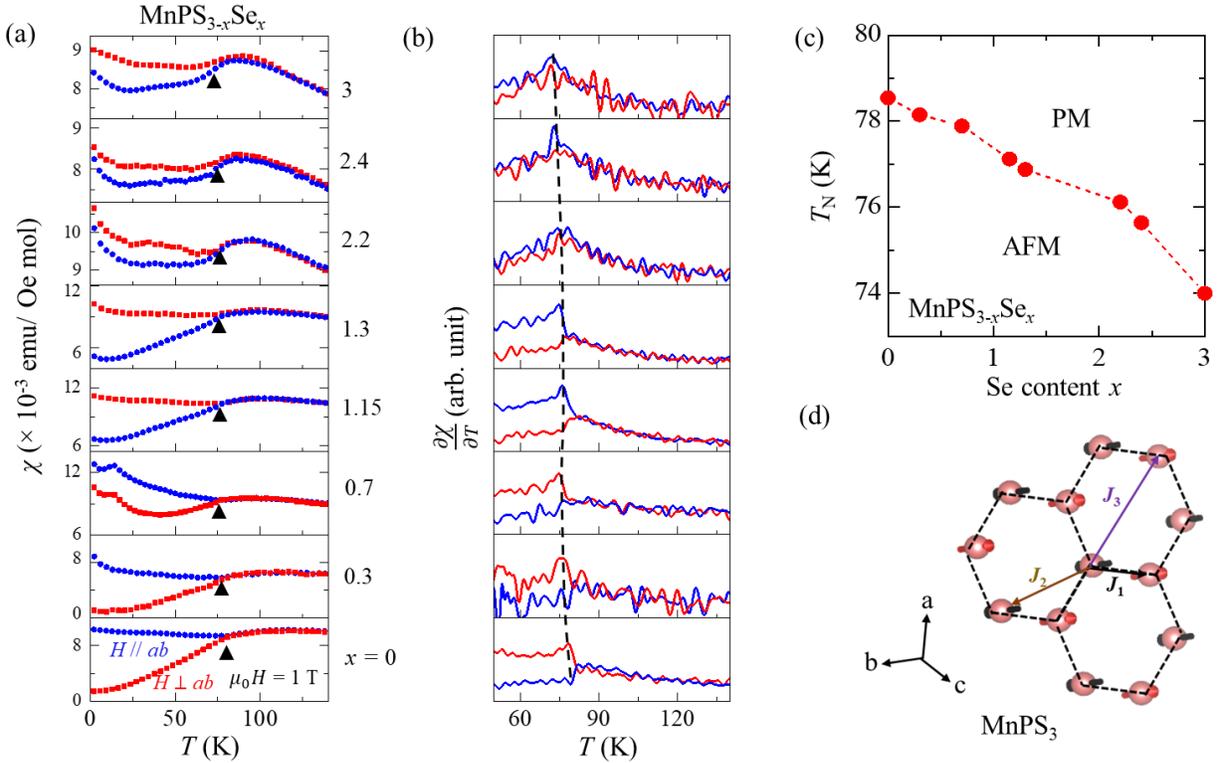

FIG. 2. (a) Temperature dependence of out-of-plane ($H\perp ab$, red) and in-plane ($H\|ab$, blue) molar susceptibility ($\chi$) of MnPS$_{3-x}$Se$_x$ ($0 \leq x \leq 3$) samples. The black triangles denote $T_N$. (b) Temperature dependence of derivative $d\chi/dT$ of MnPS$_{3-x}$Se$_x$ samples. The dashed lines denote $T_N$. (c) Doping dependence of Néel temperature ($T_N$) for MnPS$_{3-x}$Se$_x$. (d) Magnetic structure of pristine MnPS$_3$ showing nearest-neighbor ($J_1$), second nearest-neighbor ($J_2$), and third nearest-neighbor ($J_3$) interactions.

**Figure 3**

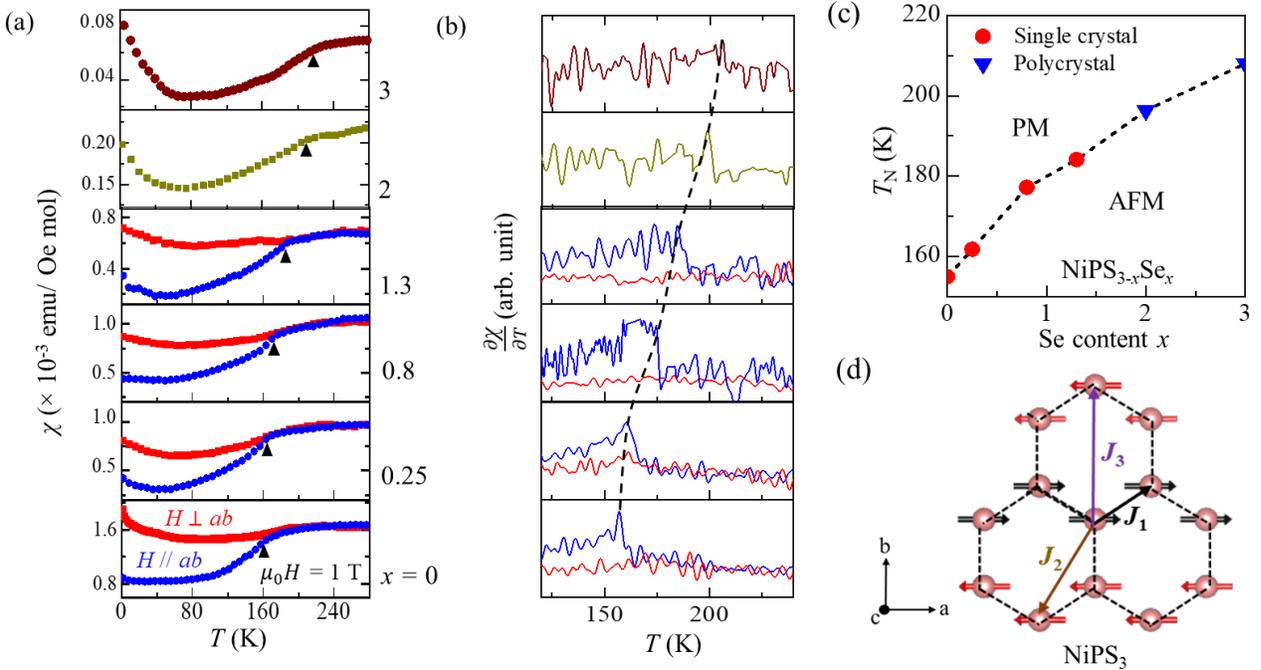

FIG. 3. (a) Temperature dependence of out-of-plane ($H\perp ab$, red) and in-plane ($H\|ab$, blue) molar susceptibility ($\chi$) of NiPS$_{3-x}$Se$_x$ ($0 \leq x \leq 3$) samples. The dark yellow and wine color data represent $x = 2$ and 3 polycrystals samples, respectively. The black triangles denote $T_N$. (b) Temperature dependence of derivative $d\chi/dT$ of NiPS$_{3-x}$Se$_x$ samples. The dashed lines denote $T_N$. (c) Doping dependence of Néel temperature ($T_N$) of NiPS$_{3-x}$Se$_x$. (d) Magnetic structure of pristine NiPS$_3$ showing nearest-neighbor ($J_1$), second nearest-neighbor ($J_2$), and third nearest-neighbor ($J_3$) interactions.

**Figure 4**

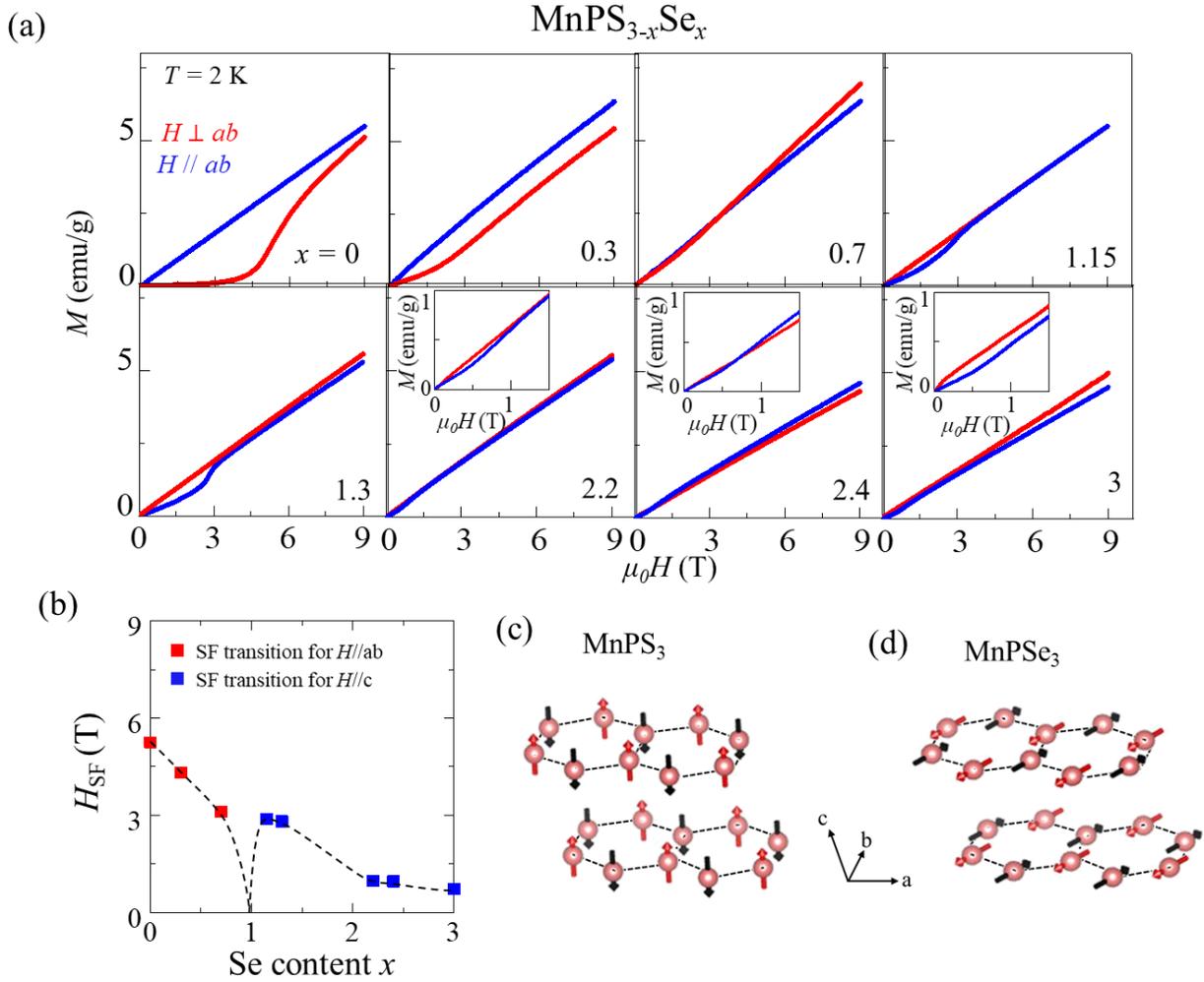

FIG. 4. (a) Field dependence of magnetization of MnPS$_{3-x}$Se$_x$ samples ($0 \leq x \leq 3$) at $T = 2$ K for out-of-plane ($H \perp ab$, red) and in-plane ($H \| ab$, blue) fields. Inset: Low-field magnetizations (b) Doping dependence of Spin-flop field ($H_{SF}$) of MnPS$_{3-x}$Se$_x$. (c) Magnetic structure of pristine MnPS$_3$. (d) Magnetic structure of pristine MnPSe$_3$.

**Figure 5**

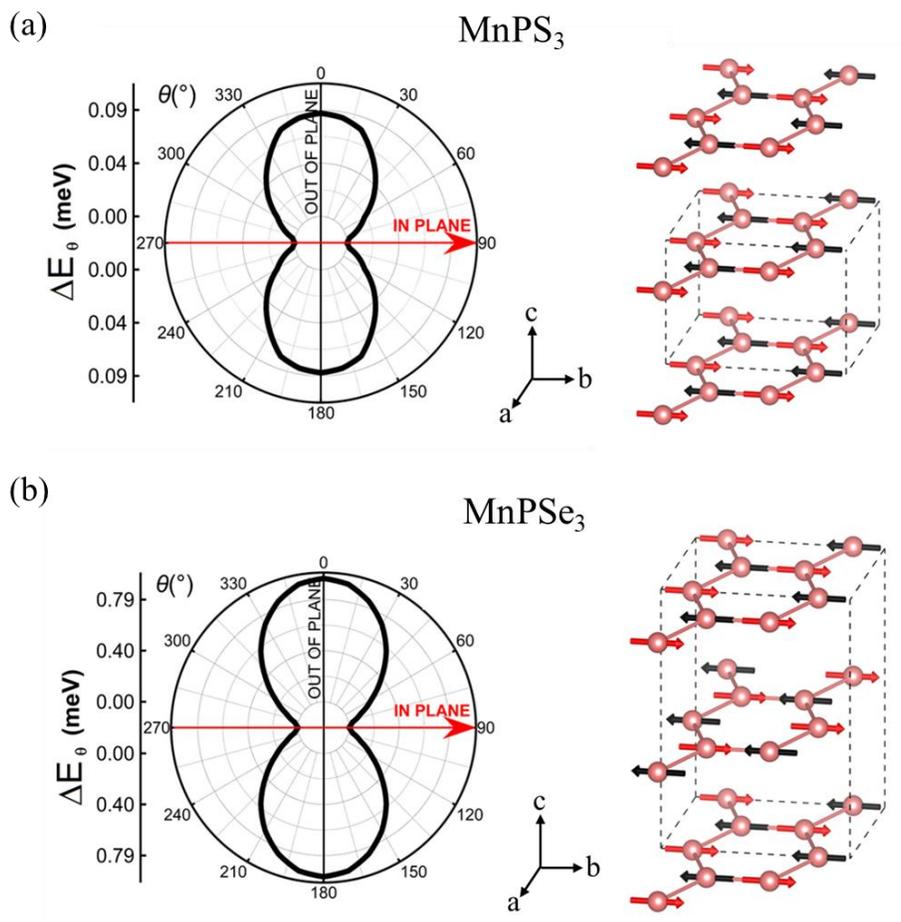

FIG. 5. (a) Left: Energy difference between the particular direction of the spins and the magnetic ground state for MnPS$_3$ system including the SOC and neglecting the magnetic dipolar interactions within the DFT studies. The latter is crucial for the proper arrangement of the spins for this system. The $\theta$ angle indicates the rotation angle from out-of-plane to in-plane position. Right: Schematic arrangement of the Mn spins in MnPS$_3$ structure. The theoretical studies predict the easy plane of magnetization in contradiction to the experimental results. (b) Left: Energy difference between the particular direction of the spins and the magnetic ground state for MnPSe$_3$ system. Right: Schematic arrangements of the Mn spins. Note, that the magnetic ordering within the layer exhibits AFM-Néel type of order, whereas the adjacent layers have antiferromagnetically aligned spins (reversed AFM-N). The easy plane of magnetization is predicted for this system.

**Figure 6**

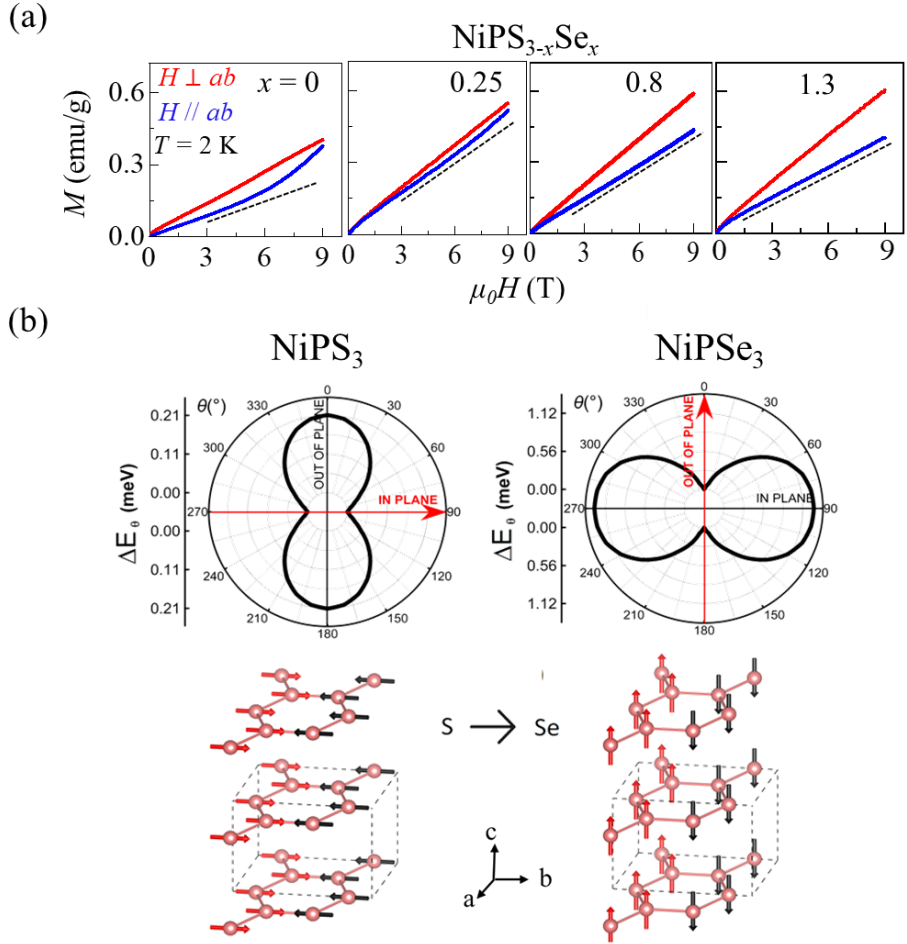

FIG. 6. (a) Field dependence of magnetization of NiPS$_{3-x}$Se$_x$ samples ($0 \leq x \leq 1.3$) at $T = 2$ K for out-of-plane ($H\perp ab$, red) and in-plane ($H\|ab$, blue) fields. The dashed lines are a guide to the eye. (b) Upper Panels: Energy difference between the particular directions of the spins and the magnetic ground state for NiPS3 and NiPSe3. The spins of the Ni atoms are collinearly aligned. The $\theta$ angle indicates the rotation angle from out-of-plane to in-plane position. The changes in rotation angle within the layer are negligible (negligible difference between the spins oriented in $a$ and $b$ directions). Lower Panels: Schematic pictures of magnetic spins for NiPS$_3$ and NiPSe$_3$. Note that, the NiPS$_3$ exhibits an easy plane (XY) of magnetization, whereas the NiPSe$_3$ has an easy axis oriented perpendicular to the layer.